\documentclass[twocolumn,aps,prb,superscriptaddress,floatfix]{revtex4}
\usepackage{graphicx}
\newcommand{\bnmr}{$\beta$-NMR}
\newcommand{\lem}{LE-$\mu$SR}
\newcommand{\Li}{$^8$Li}

\begin{document}

\title{Hyperfine Fields in an Ag/Fe Multilayer Film Investigated with
$^{8}$Li $\beta$-Detected Nuclear Magnetic Resonance}

\author{T.A.~Keeler}
\affiliation{Department of Physics and Astronomy, University of British
Columbia, Vancouver, BC, Canada V6T 1Z1}

\author{Z.~Salman}
\altaffiliation[Email address: ]{z.salman1@physics.ox.ac.uk}
\altaffiliation[Current address: ]{Clarendon Laboratory, Department of
  Physics, Oxford University, Parks Road, Oxford OX1 3PU, UK}
\affiliation{TRIUMF, 4004 Wesbrook Mall, Vancouver, Canada V6T 2A3}

\author{K.H.~Chow}
\affiliation{Department of Physics, University of Alberta, Edmonton, AB, Canada T6G 2G7}

\author{B.~Heinrich}
\affiliation{Department of Physics, Simon Fraser University,
  Vancouver, BC, Canada V5A 1S6}

\author{M.D.~Hossian}
\affiliation{Department of Physics and Astronomy, University of British
Columbia, Vancouver, BC, Canada V6T 1Z1}

\author{B.~Kardasz}
\affiliation{Department of Physics, Simon Fraser University,
  Vancouver, BC, Canada V5A 1S6}

\author{R.F.~Kiefl}
\affiliation{Department of Physics and Astronomy, University of British
Columbia, Vancouver, BC, Canada V6T 1Z1}
\affiliation{TRIUMF, 4004 Wesbrook Mall, Vancouver, Canada V6T 2A3}

\author{S.R.~Kreitzman}
\author{C.D.P.~Levy}
\affiliation{TRIUMF, 4004 Wesbrook Mall, Vancouver, Canada V6T 2A3}

\author{W.A.~MacFarlane}
\affiliation{Department of Chemistry, University of British Columbia,
  Vancouver, BC, Canada V6T 1Z1}

\author{O.~Mosendz}
\affiliation{Department of Physics, Simon Fraser University,
  Vancouver, BC, Canada V5A 1S6}

\author{T.J.~Parolin}
\affiliation{Department of Chemistry, University of British Columbia,
  Vancouver, BC, Canada V6T 1Z1}

\author{M.R.~Pearson}
\affiliation{TRIUMF, 4004 Wesbrook Mall, Vancouver, Canada V6T 2A3}

\author{D.~Wang}
\affiliation{Department of Physics and Astronomy, University of British
Columbia, Vancouver, BC, Canada V6T 1Z1}

\begin{abstract}
Low energy $\beta$-detected nuclear magnetic resonance (\bnmr) was
used to investigate the spatial dependence of the hyperfine magnetic
fields induced by Fe in the nonmagnetic Ag of an Au(40 \AA)/Ag(200
\AA)/Fe(140 \AA) (001) magnetic multilayer (MML) grown on GaAs. The
resonance lineshape in the Ag layer shows dramatic broadening compared
to intrinsic Ag. This broadening is attributed to large induced
magnetic fields in this layer by the magnetic Fe layer. We find that
the induced hyperfine field in the Ag follows a power law decay away
from the Ag/Fe interface with power $-1.93(8)$, and a field
extrapolated to $0.23(5)$ T at the interface.
\end{abstract}
\maketitle

The unique properties exhibited by thin layers of ferromagnetic metal
separated by a layer of nonmagnetic metal spacer are both interesting
and useful for applications in ``spintronic'' devices
\cite{Heinrich93AP}. In these structures the coupling between the
ferromagnetic layers oscillates between ferromagnetic (FM) and
antiferromagnetic (AF) as a function of the thickness of the
nonmagnetic spacer separating them\cite{grunburg,heinrich, parkin1,
parkin2}. This interlayer exchange coupling (IEC) is related to an
oscillating electronic spin polarization induced in the nonmagnetic
spacer due to the magnetic layers. The oscillation period ($\sim 10$
\AA)\cite{parkin3} is governed by the Fermi surface of the metal
spacer. While several theoretical models have been developed to
explain this behavior, it is difficult to establish which is the most
``correct'' since a direct measurement of the effects in the
nonmagnetic spacer is challenging. These phenomena led to the
discovery of the giant magneto-resistance (GMR) effect,\cite{footnote}
which is of great scientific interest, and also has important
technological ramifications. For example, the drastic increase in hard
disk bit density of the last 20 years was made possible by the vast
increase in sensitivity of read heads that incorporate GMR
structures. This sensitivity is a consequence of the strong field
dependence of the resistivity of GMR structures.

The most well-known model developed to explain the FM-AF coupling
oscillations in these systems is an extension of the
Ruderman-Kittel-Kasuya-Yosida (RKKY) model describing the effect of a
magnetic impurity on the conduction electrons of a nonmagnetic host
metal\cite{edwards,bruno}. The oscillations are the result of the
sharp cutoff of wave-vectors at the Fermi surface in the spacer,
resulting in an imperfect screening of the magnetic moments and
oscillations in the polarization of conduction electrons. It is via
this polarization that the magnetic layers are coupled. The period of
oscillation is thus related to the Fermi surface, with period
determined by the ``critical spanning'' wave-vectors of the spacer
material\cite{stiles}. In Ag there are two critical spanning vectors
associated with the ``neck'' and ``belly'' regions of the Fermi
surface in the (001) direction.

The small amplitude of the induced electronic polarization (due to
rapid decay away from the magnetic layer) as well as the small
physical size of typical samples (spacer thicknesses typically several
hundred {\AA} or less) makes direct measurements of the induced
polarization within the nonmagnetic spacer layer very difficult. To
date, most quantitative methods used to measure the polarization in
the spacer material are either averages of the polarization across the
entire spacer, or probe the surface of a nonmagnetic overlayer grown
on a ferromagnetic substrate. In particular it is very
difficult/impossible to directly probe the spatial dependence of the
conduction electrons polarization within the nonmagnetic layer. Such
measurements require a technique that is sensitive to the local
polarization of conduction electrons throughout the entire
spacer. M\"{o}ssbauer spectroscopy\cite{kobayashi}, perturbed angular
correlations\cite{runge}, and nuclear orientation\cite{turrell2} are
local probes, but their limited sensitivity restricts them to
measurements close the magnetic-nonmagnetic interface where the
induced fields are strongest.  In order to probe the behaviour deep
within the non-magnetic layer one requires more sensitive measurements
of the local polarization, such as low energy muons spin rotation
(\lem)\cite{luetkens2,luetkens}. The technique used herein is
depth resolved $\beta$-detected nuclear magnetic resonance (\bnmr).

In this paper we report the results of \bnmr\ measurements of the
induced hyperfine field distribution in the nonmagnetic layer of a
Ag(200 \AA)/Fe(140 \AA) magnetic multilayer (MML) prepared by
molecular beam epitaxy (MBE) on a GaAs (001) single crystal
substrate. We find that the induced hyperfine field in the Ag decays
away from the Ag/Fe interface following a power law with exponent
$\alpha=-1.93(8)$. One of the key parameters in theories describing
this effect is the exponent $\alpha$, which is both difficult to
measure or calculate. However, it has significant fundamental and
practical importance since it determines how strongly two magnetic
layers separated by a nonmagnetic spacer layer will couple.

\begin{figure}[htb]
  \includegraphics[width=\columnwidth]{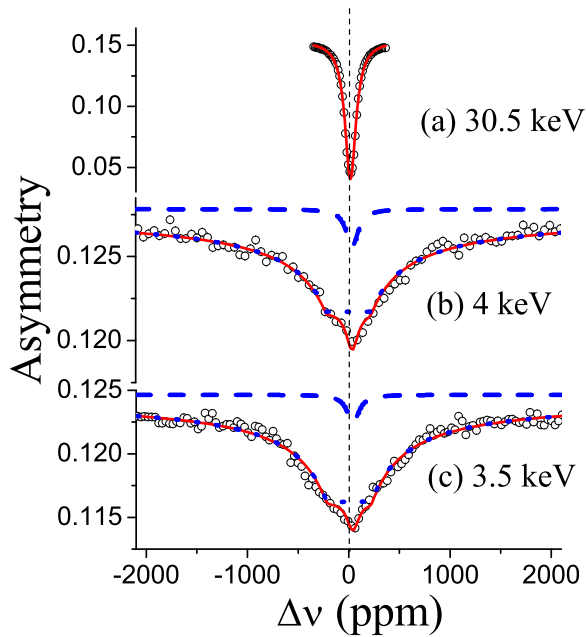}
  \caption{(color online) \bnmr\ spectra measured in the MML sample at
    room temperature and a field of 4.5 T at implantation energy (a)
    $30.5$ keV, (b) $4$ keV and (c) $3.5$ keV. The solid red lines are
    best fit to the calculated lineshapes (see text). The
    dotted/dashed lines represent the contribution from \Li\ in Ag and
    Au respectively.}
    \label{fig:fitspectra}
\end{figure}
\bnmr\ is a technique closely related to conventional nuclear magnetic
resonance (NMR). However, in \bnmr\ the signal is generated using the
$\beta$-decay properties of highly spin polarized radioactive nuclei
($\sim 10^{8}$) that have been implanted directly into the sample,
whereas conventional NMR relies on a much larger number ($\sim
10^{18}$) of intrinsic nuclei to generate a signal. \bnmr\ experiments
conducted in the ISAC facility at TRIUMF use a beam of spin polarized
radioactive $^{8}$Li ($I=2$,$\gamma = 6301.5$ kHz/T, $\tau=1.21$ s)
which is implanted into the sample and used as a spin probe. The
\Li\ nuclear polarization, which is the quantity of interest, is
monitored through its beta decay where a high energy electron is
emitted preferentially opposite to the direction of its nuclear
spin. The polarization is measured as a function of the frequency of
an applied transverse radio frequency (RF) field. The resulting
resonance is a sensitive probe of the local internal electronic and
magnetic environment. The implantation energy can be adjusted in the
range $0.1-30.5$ keV, corresponding to mean implantation depths of
between $2$ and $200$ nm from the sample surface. Previous studies on
an Au(40 \AA)/Ag(800 \AA)/Fe(14 \AA) (001) film demonstrated our
ability to make depth resolved \bnmr\ measurements in the different
layers, as well as our sensitivity to the induced hyperfine fields in
the Ag close to the Fe\cite{keeler}. This ability to extract
information about the local magnetic environment as a function of
depth on a nm length scale distinguishes low energy \bnmr\ from
conventional $\mu$SR and NMR. It is similar to \lem\ in this respect
but there are significant differences, e.g. the different time scale,
so that the two methods are often complementary. A more complete
description of the \bnmr\ technique can be found
elsewhere\cite{bnmr,morris}.

The Au(40 \AA)/Ag(200 \AA)/Fe(140 \AA) MML sample was grown using MBE
on a GaAs (001) single crystal substrate.  The substrate was sputtered
clean and annealed to yield large flat terraces on which a 140 \AA\ Fe
layer was grown. The small lattice mismatch between GaAs and Fe (-1.4
$\%$) allows growth of body centered cubic (bcc) Fe (001) into well
ordered layers. Then a 200 \AA\ layer of face centered cubic (fcc) Ag
was grown on Fe following the (001) orientation with its lattice
rotated by $\pi/4$.\cite{etienne} The sample was finally capped with a
protective 40 \AA\ Au layer. The thicknesses of the layers were
monitored during growth using a calibrated quartz crystal
microbalance.

\begin{figure}[h]
  \begin{center}
    \includegraphics[width=\columnwidth]{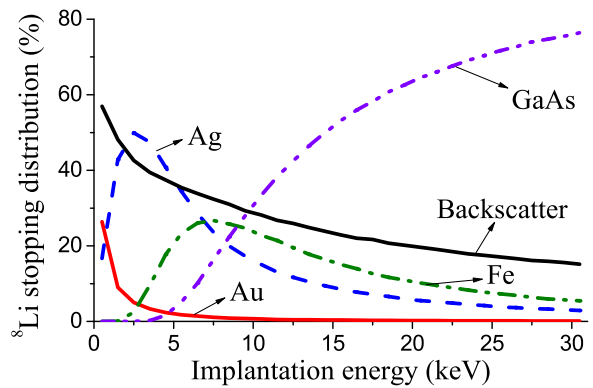}
    \caption{(color online) Percentage of \Li\ stopping in each layer
      of the MML sample as a function of implantation energy,
      calculated using TRIM.SP Monte-Carlo simulation.}
      \label{fig:TRIM2}
  \end{center}
\end{figure}
The sample was placed in the \bnmr\ spectrometer in an applied
magnetic field of $4.5$ T normal to the film surface. Representative
\bnmr\ spectra at room temperature and 3 different implantation
energies are shown in Fig.~\ref{fig:fitspectra}. At full implantation
energy ($30.5$ keV) most of the \Li\ is implanted into the GaAs
substrate. The resonance in Fig.~\ref{fig:fitspectra}(a) fits well to
a Lorentzian lineshape with a width of $4$ kHz, as expected for \Li\
in GaAs.\cite{chow} Figs~\ref{fig:fitspectra}(b) and (c) show the
spectra obtained with implantation energies $4.0$ and $3.5$ keV
respectively. As shown in Fig.~\ref{fig:TRIM2}, at these energies
TRIM.SP Monte-Carlo simulations \cite{eckstein} predict that most of
the \Li\ stops in the Ag layer ($\sim 50 \%$), with a small amount
($\sim 3 \%$) stopping in the Au layer. The remaining \Li\ is
backscattered ($\sim 37 \%$) or stops in the Fe ($ \sim 10 \%$). Note,
since backscattered \Li\ stops outside the RF coil they do not
contribute to the measured resonance line. Similarly, \Li\ stopping in
Fe experiences a very large magnetic field and therefore produces a
resonance outside our frequency window \cite{Matsuta01HI}. Recent
\bnmr\ measurements show that the intrinsic \Li\ resonance linewidth
in a thin Ag film is $0.5-1$ kHz.\cite{morris} In contrast, the
linewidth observed in Fig.~\ref{fig:fitspectra}(b) and (c) is an order
of magnitude larger. This broadening is attributed to the induced
hyperfine magnetic field in Ag due to the Fe magnetic layer.

\begin{figure}[htb]
  \includegraphics[width=\columnwidth]{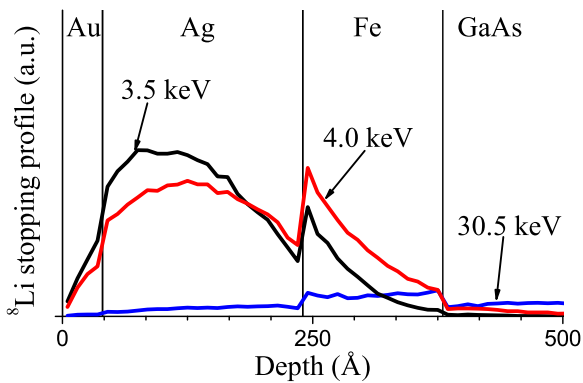}
  \caption{(color online) The \Li\ stopping profile in the MML sample
    generated using TRIM.SP Monte-Carlo simulations for implantation
    energies of 3.5, 4.0 and 30.5 keV. At lower energies \Li\ stops
    predominantly in the Ag layer, while at the full energy nearly all
    \Li\ stops in the GaAs substrate.}
    \label{fig:TRIM}
\end{figure}

\begin{figure}[h]
  \begin{center}
    \includegraphics[width=\columnwidth]{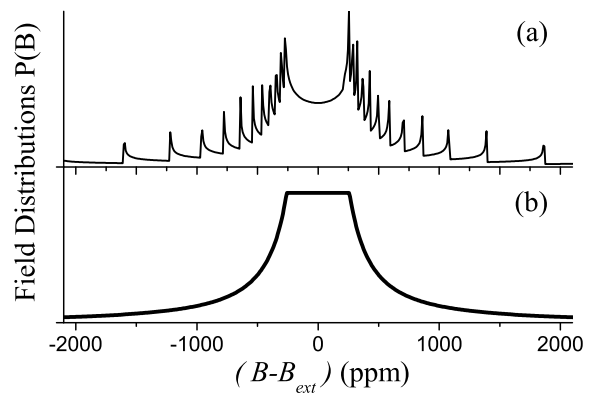}
    \caption{Simulated lineshape, in an applied field $B_{ext}$,
      generated for induced hyperfine fields that follows (a) an
      oscillating distribution described by Eq.~(\ref{RKKYB}) and (b)
      a uniform distribution within the envelope described by
      Eq(\ref{Power}). Both lineshapes were calculated using
      $\lambda=11.7$ \AA, $\alpha=-1.9$, $B_0=0.23$ T and
      $B_{ext}=4.5$ T.}
      \label{fig:modellines}
   \end{center}
\end{figure}

We now discuss the modelling of the lineshapes obtained, such as those
shown in Fig.~\ref{fig:fitspectra}. Following the RKKY-based
theoretical description \cite{Bruno91PRL}, the induced hyperfine field
in Ag as a function of the distance $(x)$ from an ideal Ag/Fe
interface, follows
\begin{equation} \label{RKKYB}
  B(x)=\sum_{i=0}^1 B_i \left( \frac{x}{\lambda_i} \right)^{\alpha_i}
    \sin \left( \frac{2\pi x}{\lambda_i}+\phi_i \right),
\end{equation} 
where $\lambda_i=2 \pi/k_i$ are the Fermi wavelengths associated with
the two critical spanning vectors, i.e. the belly and neck. The
expected resonance lineshape, representing the magnetic field
probability distribution for the implanted \Li, is calculated using
Eq.~(\ref{RKKYB}), and the stopping \Li\ profile in the Ag layer
determined using TRIM.SP calculations (Fig.~\ref{fig:TRIM}).
An example calculated using Eq.~(\ref{RKKYB}), with a contribution
only from the belly spanning vector ($\lambda=11.7$ \AA,
$\alpha=-1.9$, $B_0=0.23$ T and $B_{ext}=4.5$ T.), is shown in
Fig~\ref{fig:modellines}(a). The contribution from the neck spanning
vector is negligible, since its associated wave length ($\sim 4.85$
\AA) is on the same length scale of the distance between neighboring
implanted \Li, and hence its contribution is averaged out over the
\Li\ stopping sites in the Ag lattice. The distinguishing features of
this lineshape are the peaks on either side of the applied field
($B_{ext}$) resulting from Van Hove singularities. The large inner
double peaks result from the non-zero hyperfine fields at the Ag
farthest from the Ag/Fe interface. However, it has been shown that
even slight interface roughness tends to wipe out the short wavelength
oscillations in the electron polarization\cite{wang} since the
distance to the interface no longer has a well defined value over
lateral distances larger than the terrace width. Furthermore a
vertical mismatch between atomic planes, as little as $0.8 \%$ for
Ag/Fe(001) also leads to suppression of both the long and short
wavelengths.\cite{celinski} The \Li\ beam averages over the area of
the beam spot ($\sim 3$ mm diameter), therefore we do not expect to
observe oscillations of the induced hyperfine fields. This will have
the effect of ``smearing'' out the oscillations in the induced field
[Eq.~(\ref{RKKYB})]. Therefore we use a phenomenological form for the
induced field distribution:
\begin{equation} \label{Power}
B_{max}(x)=\frac{B_0}{1+(\lambda_F / x)^\alpha}
\end{equation}
with $\lambda_F = 2 \pi /k_F$ taken as the long period Fermi
wavelength of Ag ($11.7$ \AA). We assume that at a particular
distance, $x$, from the interface, the hyperfine field is equally
likely\cite{footnote2} to have any value between
$B_{ext}{\pm}B_{max}$. This form has several advantages: 1) it gives a
value of $B_0$ for the hyperfine field right at the Ag/Fe interface,
2) it maintains the asymptotic power law behaviour, $x^{\alpha}$,
predicted by RKKY, 3) it avoids the unphysical divergent behaviour at
the Ag/Fe interface in Eq.~(\ref{RKKYB}). The lineshape, shown in
Fig~\ref{fig:modellines}(b), results from this form of field
distribution. Note that it does not have the peaks associated with the
oscillating magnetic field, but it is symmetric and exhibits a
characteristic ``flat top''. This low field cutoff originates from the
fact that the hyperfine field does not decay to zero at the Ag/Au
interface. This lineshape is consistent with the \bnmr\ spectra in
Figs~\ref{fig:fitspectra}(a) and (b).

Note that, in contrast to the calculation above and the lineshape in
Fig.~\ref{fig:modellines}(b), the experimental \bnmr\ spectra show a
sharp peak and not a ``flat top'' feature. This is attributed to the
small amount of \Li\ that stops in the thin Au capping layer (dashed
lines in Fig.~\ref{fig:modellines}(b) and (c)). These spectra were fit
to the sum of our model lineshape, Fig~\ref{fig:modellines}(b), and a
Lorentzian to account for the small signal from Au. The contribution
to the measured resonances from Ag and Au, as obtained from the fit,
are represented by the dotted and dashed lines in
Fig.~\ref{fig:fitspectra} respectively. The induced field parameters
extracted from the fits in Fig.~\ref{fig:fitspectra}(b) and (c) are
$\alpha=-1.93(8)$ and $B_0=0.23(5)$ T (common for both spectra). The
Lorentzian fit for the resonance in Au was found to be $\sim 4$ kHz
wide, and shifted approximately $+40$ parts per million (ppm) relative
to the GaAs reference. The Knight shift of the Au is comparable with
previous measurements of intrinsic Au ($+60(20)$ ppm)
\cite{macfarlane}, but the width is about twice that of previous
measurements. This may be due to the higher RF power used in these
experiments but could also be due to the induced fields from Fe
extending into the Au layer. In addition, we find that the ratio of
the contribution from Au to Ag from the fit parameters (i.e. the ratio
of the areas of the resonance in Au to that in Ag) is $\sim 3\%$, in
reasonable agreement with $\sim 6\%$ from TRIM.SP calculations
(Fig.~\ref{fig:TRIM2}). The $3\%$ difference may be due to the limited
accuracy of TRIM.SP in predicting ion implantation profiles especially
at low implantation energies.\cite{eckstein,Morenzoni02NIMB}

The extracted value of $B_0=0.23(5)$ T, which is the hyperfine
coupling of $^8$Li at the Ag/Fe interface, is in reasonable agreement
with the calculated value $0.30$ T for the induced hyperfine field at
the first Ag layer at the interface.\cite{rodriguez} Also the
asymptotic power of the induced field, $\alpha=-1.93(8)$, agrees very
well with the theoretical value $-2$ predicted by RKKY
theory.\cite{Yafet87PRB,wang,Bruno91PRL} It is however larger than the
values $\alpha=-0.4(1)$ and $-0.8(1)$ measured using \lem\ in Fe(40
\AA)/Ag(3000 \AA)(001) \cite{luetkens2} and Fe(40 \AA)/Ag(200
\AA)/Fe(40 \AA)(001) \cite{luetkens} samples, respectively. Similarly,
using Cu NMR to measure the spin polarization profile in multilayers
of Ni/Cu, Goto {\em et al.} \cite{Goto93JPSJ} found $\alpha =-1$. In
all those measurements, induced fields of the form Eq.(\ref{RKKYB})
were assumed and the magnetic field parallel to the surface was used
to perform these measurements, while in our measurements the field was
applied perpendicular to the surface. In principle, this should not
affect the value of $\alpha$ \cite{Goto93JPSJ}. The source of
discrepancy between our results and those from
Refs. \cite{luetkens2,luetkens,Goto93JPSJ} may be the reduced
sensitivity of the previous measurements to the interface region. In
the \lem\ measurements, the contribution of the 3 nm region of the
Ag/Fe interface is negligible \cite{luetkens2}, since the muons in
this region experience a field which is too high to be measured. Note
that both NMR and \lem\ measurements are performed in the time domain,
i.e. the high field contribution occurs at early times, therefore the
dead time associated with the measurement decreases the sensitivity to
high field regions (interface). In contrast our measurements are
performed directly in the frequency domain and therefore have no such
effect (provided that one sweeps a sufficiently large frequency
range). In addition, in NMR measurements it is extremely hard to
account for the contribution of all nuclei in the spacer since the
resonance cannot be normalized, while the method of detection used in
\bnmr\ enables detection of the signal from all implanted spin probe
nuclei. Finally, we would like to point out that calculation based on
the quantum-interference model \cite{Ishiji06PRB} predict an
oscillating polarization that involved three terms with $\alpha=-1,-2$
and $-3$. Thus, it may be that the different techniques are sensitive
to different terms.

In conclusion, we have carried out depth resolved low energy \Li\
\bnmr\ to measure directly the hyperfine field profile in an Ag layer
induced by a magnetic Fe layer. No indication of an oscillating
hyperfine field is observed. However, we find that the induced fields
decrease away from the Ag/Fe interface following an asymptotic power
law $x^{- 1.93(8)}$ in good agreement with theoretical calculations
based on RKKY theory. The induced field at the Ag/Fe interface
$B_0=0.23(5)$ T is also in good agreement with calculations.

This research was supported by the Center for Materials and Molecular
Research at TRIUMF, the Natural Sciences and Engineering Research
Council of Canada and the Canadian Institute for Advanced Research. We
would like to thank W. Eckstein from MPI fur Plasmaphysik in Garching,
Germany, who developed the TRIM.SP code we used to generate the
stopping profiles. We would especially like to thank Rahim Abasalti,
Bassam Hitti, and Donald Arseneau for their expert technical support.


\begin{thebibliography}{xx}
\bibitem{footnote} IEC is generally not a necessary condition for
  observing the GMR effect. However, if one does not have a spin-valve
  structure, an antiferromagnetic coupling is necessary to observe the
  GMR effect.
\bibitem{footnote2} Models where increased weight is given to the
extrema regions ($B_{ext} \pm B_{max}$) were also considered. However,
the fitting parameters and lineshapes produced where very similar to
those obtained using the simpler model. Therefore we used the simpler
model to analyse our results.
\end{thebibliography}
\end{document}